# Beyond Ternary OPV: High-Throughput Experimentation and Self-Driving Laboratories Optimize Multi-Component Systems

Stefan Langner*[a], Florian Häse*[b,c,d,e], José Darío Perea[a], Tobias Stubhan[f], Jens Hauch[f], Loïc M. Roch[b,c,d,e], Thomas Heumueller[a], Alán Aspuru-Guzik[c,d,e,g], and Christoph J. Brabec[a,b]

[a] Institute of Materials for Electronics and Energy Technology (i-MEET), Department of Materials Science and Engineering, Friedrich-Alexander University Erlangen-Nürnberg, Martensstrasse 7, 91058 Erlangen, Germany
[b] Department of Chemistry and Chemical Biology, Harvard University, Cambridge, MA 02138, USA
[c] Department of Chemistry, University of Toronto, Toronto, ON M5S 3H6, Canada
[d] Department of Computer Science, University of Toronto, Toronto, ON M5S 3H6, Canada
[e] Vector Institute for Artificial Intelligence, Toronto, ON M5S 1M1, Canada
[f] Forschungszentrum Jülich GmbH, Helmholtz-Institute Erlangen-Nürnberg for Renewable Energies, Immerwahrstr. 2, 91058 Erlangen, Germany
[g] Canadian Institute for Advanced Research (CIFAR) Lebovic Fellow, Toronto, ON M5S 1M1, Canada

Fundamental advances to increase the efficiency as well as stability of organic photovoltaics (OPVs) are achieved by designing ternary blends which represents a clear trend towards multi-component active layer blends. We report the development of high-throughput and autonomous experimentation methods for the effective optimization of multi-component polymer blends for OPVs. A method for automated film formation enabling the fabrication of up to 6048 films per day is introduced. Equipping this automated experimentation platform with a Bayesian optimization, a self-driving laboratory is constructed that autonomously evaluates measurements to design and execute the next experiments. To demonstrate the potential of these methods, a four- dimensional parameter space of quaternary OPV blends is mapped and optimized for photo-stability. While with conventional approaches roughly 100 mg of material would be necessary, the robot based platform can screen 2,000 combinations with less than 10 mg and machine learning enabled autonomous experimentation identifies the stable compositions with less than 1 mg.

**Introduction**

With the development of novel non-fullerene acceptors fundamental performance limitations of fullerene based organic solar cells (OSC) have been overcome.[1] In the currently highest performing system PM6:Y6 ternary additives are used to improve charge carrier lifetime and mobility resulting in significantly improved performance with efficiencies up to 16.5%.[2,3] Typically in ternary systems the host D:A ratio is kept constant, while only the content of additive is varied due to limited experimental resources. The existence of other optima beyond these constraints is usually not investigated. On the other hand, also for improvements of stability, ternary additives have successfully been introduced to stabilize morphology or prevent oxidation. These desired effects of additives are typically reverted into detrimental effects depending on the concentration and compatibility of the additive with the host system. Finding a fine balance is necessary to determine if a given additive can lead to a performance enhancement for a given host system. Currently the Achilles heel of the highest performing organic photovoltaic (OPV) systems is device stability, which motivates the use of a fourth stabilizing additive in performance optimized ternary systems. On the pathway to such highly complex optimization problems with multi-dimensional composition space and hundreds of possible candidates for performance enhancing additives, novel experimental methods are necessary.

High-throughput experimentation (HTE) addresses these challenges by assisting the researcher in material synthesis, sample processing and characterization. High throughput methods for polymer samples have been used for adhesion evaluations [4] and cell biology research.[5,6] One of the main challenges of combinatorial research in OPV is the scalable fabrication of high-quality individual films. Recent approaches are based on gradient coating or ink-jet printing, which are usually limited by the number of mixable components, the number of samples or the need for specific ink properties.[7–9] An HTE coating method for organic semiconductors with a scalable number of mixable components and scalable number of films is not yet established. Therefore, we introduce a robot-based production of OPV-films via drop-casting that can mix a large number of components to form up to 6,048 films of different compositions per day.

Applying this HTE system allows to screen such a vast number of compositions containing hundreds of possible components that conventional experimental planning and evaluation capabilities are exceeded. Therefore, as a second approach, we combine our HTE production line with a machine learning (ML) tool,[10–13] to form a self-driving laboratory (SDL), for accelerated process optimization and materials discovery.

In this ML-driven approach, new experiments are suggested based on all previously collected measurements. To this end, the ML decision-maker infers the outcomes of all possible experiments, leveraging statistical correlations identified from the prior measurements, and suggests the most informative ones for future evaluation.

To the best of our knowledge, the first self-driving laboratory reported an autonomous chemical synthesis where all the steps of the experimental procedure are controlled by a computer and driven by a local optimization algorithm.[14] Since then, the self-driving approach was used to optimize carbon nanotube growth,[15] polyoxometalate clusters, metal-based alloys,[16] and organic synthesis reactions[17,18]. More recently, in an area related to this work, a thin-film self-driving robot was used to improve film quality and thermal stability of hole-transport materials for clean energy technologies.[19]

Herein we demonstrate the benefits of HTE and the self-driving approach on the design of photostable material composites for OPV. While recently lifetimes of up to 10 years under continuous illumination in nitrogen atmosphere were reported,[20] photo-stability of OSC materials under the influence of oxygen and moisture is a critical challenge. Currently available encapsulation films for OPV materials are a major cost factor of the final product while their water vapor transmission rate is typically $10^{-3}$ g/(m$^2$day). As higher quality barrier films are currently not manufacturable at a competitive price, optimizing active layer materials for photo-stability in the presence of oxygen and water is a critical step for commercial applications. Interestingly, complex photo-chemical interactions between donor, acceptor and oxygen can suppress or enhance photo-oxidation in D:A blends.[21,22] Further, adding a third component to the active layer may increase thermal and light stability.[23,24] However, the degradation behaviors of higher dimensional multi-component systems may depend on various interactions between the components.

Therefore, we use an optimization of four component active layer blends as an experimental proof of concept for our high throughput film formation and characterization system with machine learning enabled self-driving capability. In the following, grid based high-throughput experimentation will be compared to the self-driving approach (**Figure 1C**).

**High-throughput Experimentation**

The automated platform (**Figure 1B and Figure S1**) employed for this study was already used in our group to successfully synthesize and characterize organo metal-halide perovskites and organic nanoparticles with outstanding precision.[25,26] To form high quality drop-cast films on inert carriers, we optimized a system to create separate wells on a glass plate. Printing a well structure with UV-curable epoxy allowed a minimal height of the separating walls to reduce material accumulation at the walls resulting in high quality films with a large homogenous area (**Figure 1C and Figure S2A**). Demonstrating the scalability of our approach, two quaternary blend systems were investigated simultaneously. The first system contains PTB7-Th and P3HT as polymer donors (white squares in **Figure 1A**), while in the second system PBQ-QF is used (black squares in **Figure 1A**). In both systems, PCBM and oIDTBR are employed as acceptors. Overall for both material systems, almost 2,100 single films were fabricated and tested within 7 days with more than 5,000 recorded absorbance spectra, see experimental section and **Figure S2B**.

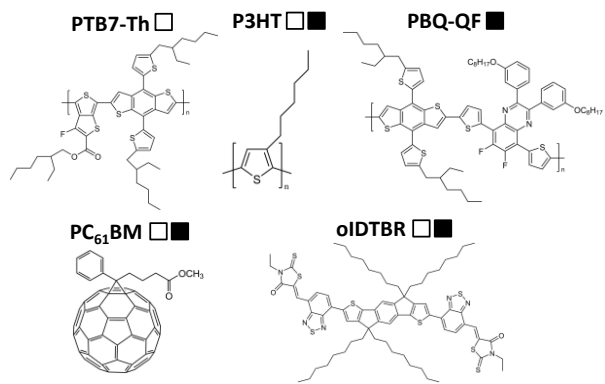
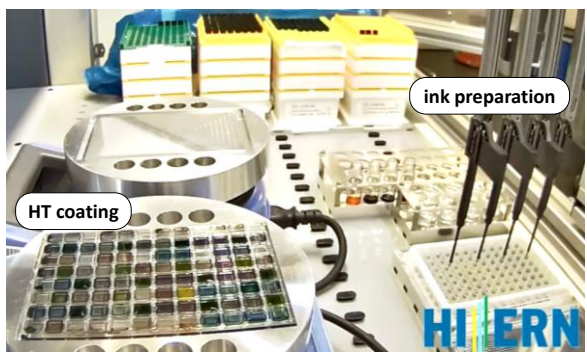
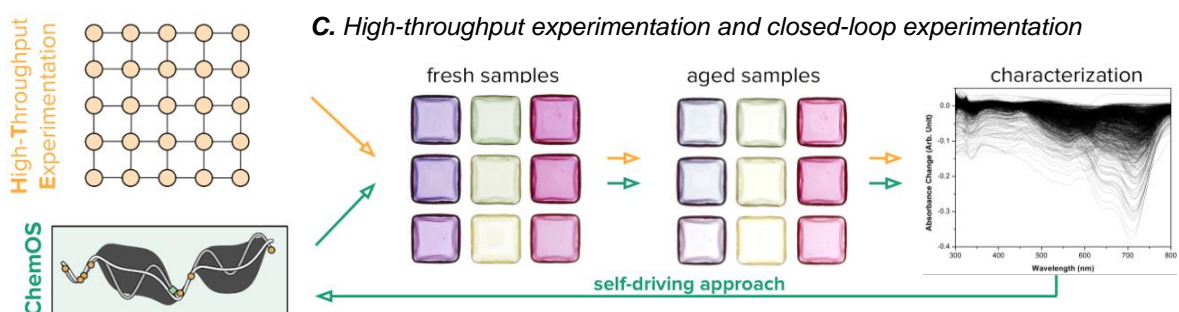

**Figure 1.** A) Representations of the three polymer donors and the two small molecule acceptors. Note that the first quaternary system consists of P3HT, PBQ-QF, PCBM, and oIDTBr (■) and the second of P3HT, PTB7-Th, PCBM, and oIDTBr (□). B) Side view of the automated platform for ink formulation, coating and characterization. C) Experimental workflow with the two approaches adopted in this study: conventional high-throughput experimentation via grid and the self-driving approach with the ChemOS software package.

In **Figure 1C**, the experimental workflow of the robot process is illustrated schematically. Starting with a pre-defined grid of compositions (or the ML-self-driving approach, see subsection 3), individual inks are formulated automatically using a commercially available liquid-handling robot. The inks are drop-cast onto customized 96-well glass substrates to form uniform semiconductor layers. The glass plates are then illuminated with metal halide lamps at one sun intensity for 18 hours in ambient air to induce photo-oxidation of the active layers, which can be measured by absorbance loss. Absorbance spectra were recorded before and after light treatment using a microplate reader.

The reproducibility of film formation and degradation is very high with an $R^2$ of 0.97 for 309 tested samples **(Figure S3)** while the film thickness, as measured by a profilometer, ranged between 70-80 nm. Given the accuracy of our dispensing system, we allowed variations of all four components (PTB7-Th, P3HT, oIDTBR, PCBM) in two percent steps resulting in 23,426 possible compositions. For the first high throughput test, a subset of 1,041 points was selected containing all single components, binary and ternary variations with 10 wt.-% increment (202 samples) as well as 820 randomly selected quaternary samples plus 19 duplicates for reproducibility verification. For the second system, the same set of compositions was used while only PTB7-Th was replaced by PBQ-QF.

The degradation behavior of the quaternary system containing PTB7-Th is depicted in **Figure 2A** (left side) where degradation is denoted as the integral change of the absorption spectra for each composition before and after degradation. It clearly shows strong degradation of PTB7-Th with a relative absorbance loss of 68%, while P3HT lost around 19%. The two acceptor materials exhibited minimal change in absorbance. Moreover, it is observed that the binary series of PTB7-Th toward PCBM or oIDTBR (see also **Figure S4 and S5)** leads to a linear increase of photo-stability with acceptor content. But non-linear trends are observed as well, as already slight amounts of P3HT (~10 wt.-%) stabilize PTB7-Th completely leading to blend stabilities similar to pristine P3HT. More surprisingly, blends containing both oIDTBR and PCBM show a drastic destabilization effect causing a large instable area in the quaternary space (Figure 2A). As this is also observed in a binary mixture of PCBM and oIDTBR (absorbance loss up to 74%) it seems to be caused by a specific interaction between the two acceptors that may be rooted in the formation of a specific morphology that might facilitate the movement of oxygen molecules through the active layer. In the second quaternary system, where PTB7-Th is replaced with PBQ-QF, the same PCBM:oIDTBR instability area is discovered, while most other compositions show improved photo-stabilities, which is in line with the high stability of pristine PBQ-QF **(Figure S4-S6)**. A detailed viewing of the robot process and the 4D stability-space is found as cinematic illustration in the supporting information.

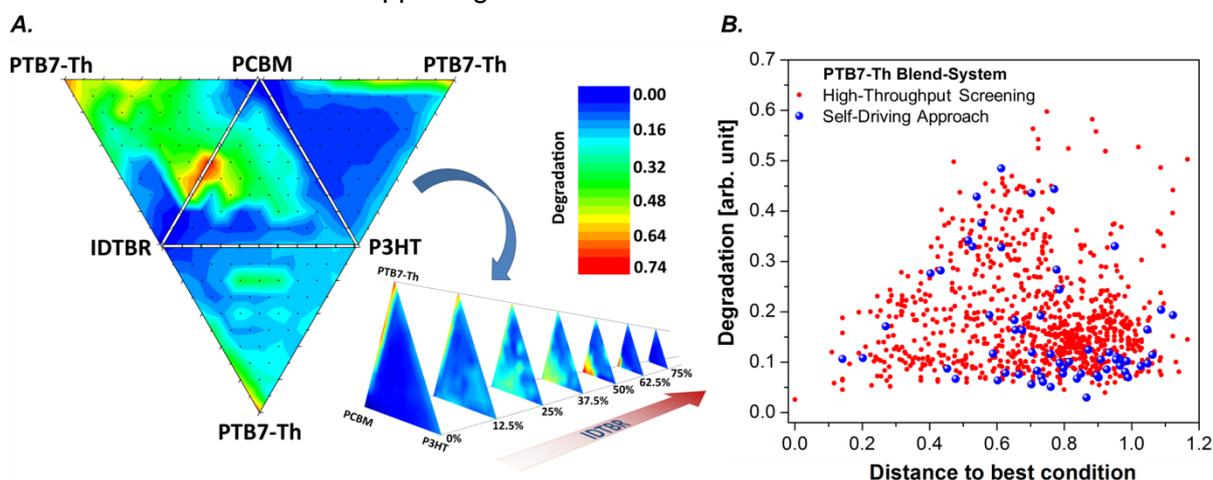

*Figure 2.* A) Photo-stability of the quaternary system containing PTB7-Th illustrated by the 4D hypersurface and pyramid cross sections. Degradation is defined as the relative spectral loss in absorbance. Black dots represent measured data points, while the color map is interpolated. B) Comparison the covered experimental space between grid-based high-throughput experimentation and self-driving optimization. The compositional distance to the most stable blend is calculated as $distance = \sqrt{\Delta x^2 + \Delta y^2 + \Delta z^2 + \Delta t^2}$) with x,y,z,t representing the weight fractions of the four blend components.

**Self-driving Laboratory**

Instead of a large predefined grid, we utilized intelligent experimental design for the self-driving approach (SDA). The machine learning driven software package ChemOS coordinates the flow of information between the automated equipment, the researchers and the experiment planning module.[27,28] The software enables the remote exchange of experimental parameters and measurements, which enables the operation of experiment planning strategies and robotics platforms at different locations. In line with the identified steps of closed-loop experimentation,[10] we extended ChemOS by an additional module which maps the parameter space of the experiment planner to technical constraints of the robot (see Section S.1).

From the experiment planner, we used the active learning global optimizer Phoenics to learn and navigate the multi-dimensional parameter space. This ML strategy learns by doing and does not need to be trained with prior measurements. Following a Bayesian optimization strategy, a surrogate model is constructed from parameter kernel densities estimated via Bayesian neural networks. This architecture enables a favorable linear scaling and reduces the computational cost of the optimization algorithm in contrast to, for examples, approaches based on Gaussian processes. Phoenics also streamlines the throughput of automated solutions via a sampling parameter, which explicitly controls the sampling behavior of the algorithm, gradually exploring and exploiting. Alternating the sampling behavior during the experimental campaign has been shown to accelerate the optimization process and reduce the number of samples.[29] In this experiment, ChemOS leveraged Phoenics to suggest four blends per closed-loop event, and increases the experimentation throughput by simultaneously coordinating the self-driving approach for the two studied blend systems. To ensure that any active layer composition which will be found contains an OPV relevant donor:acceptor ratio, a boundary condition limits the D:A ratio from 1:5 to 5:1. Starting from 4 random compositions and their stability results, ChemOS iteratively suggested the next set of 4 compositions which were fabricated, degraded and characterized. As an initial test run 15 iterations requiring 15 consecutive 18 h degradation tests were made. Here, two identical samples were fabricated for each composition to probe experimental noise.

While the SDA focused more on the stable regions in an exploitative search toward a global minimum an explorative component probed numerous stable and unstable compositions covering largely the same experimental space as the high throughput test. The experimental space for both HTE and SDA in the PTB7-Th based system is displayed in **Figure 2B** (for PBQ-QF see **Figure S7**). Here, each data point represents the stability of a quaternary composition plotted over their distance (in terms of compositional difference) to the most stable composition as found by grid-HTE. The results demonstrate how the Bayesian optimization reconstructed the stability distribution of grid-HTE with a much smaller number of samples (only 7%) compared to HTE. Generally, photo-stable compounds can be found along the entire distance axis, indicating a broad area of stable compositions rather than a single point minimum (**Figure S8**). To accommodate for such broad minima, the merit function had been adjusted after an initial test run (see SI). It is observed that the self-driving approach is able to find competitive photostable blends within only 15 learning iterations (4 samples each). The best compositions found by the two formulated merit functions are on a par with the best compositions screened by grid-HTE (**Table S1**).

**Statistical Comparison**

To have a large number of tests that allow statistical comparisons between HTE and SDA, virtual experiments were performed on a calculated stability grid. From the data collected for the two blend systems during the HTE runs, we constructed probabilistic regression models to emulate the response surfaces of the two experiments (Figure 2A and Figure S5). Then models, or virtual robots, were set up as Bayesian neural networks (BNNs) and were trained to predict the photo-stability for any possible set of material compositions for the two blend systems. This approach has already been demonstrated in the context of emulating HPLC calibrations.[30] Details on the network architectures, the training procedures and the prediction accuracies on the observed (training sets) and unobserved data (test sets) for both blends are reported in the supporting information (see Section S.3). The virtual robots are made available on GitHub.[31]

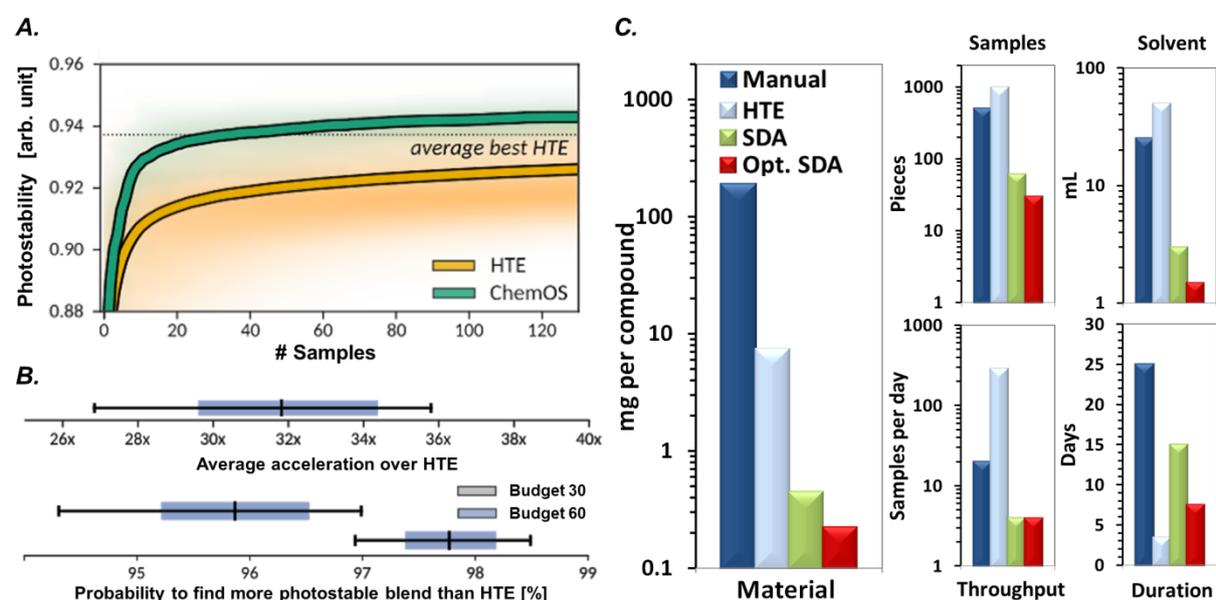

*Figure 3.* A) Performances of high-throughput experimentation (red traces) and ChemOS/Phoenics (blue traces) on the virtual robots related to the PTB7-Th based system. Note that the traces were averaged over 10,000 independent grid samplings and 200 independent ChemOS optimizations. B) Top: Statistical output of the virtual robot showing the acceleration of SDA over HTE; Bottom: Confidence to improve on HTE within given budget of 30 or 60 samples. Both are related to the PTB7-Th based system. C) Consumption comparison between manual testing, high-throughput experimentation, SDA with budget of 60 samples and virtually optimized SDA with budget of 30 samples. The calculations are limited to one quaternary system.

Our results on the virtual robots (**Figure 3A**) indicate that the self-driving approach requires on average 27 samples on the PTB7-Th blend system to identify a blend which is at least as photostable as the most stable blend discovered by grid-HTE. Furthermore, we find that within a given budget of 30 (60) samples, the SDA identifies more photostable blends than HTE with a chance of about 96% (97.5%) (**Figure 3B**). Based on these results the discovery of photostable blends is accelerated by a factor of *ca.* 33x for PTB7-Th (and *ca.* 32x for PBQ-QF, **Figure S9**) over conventional HTE as indicated by the virtual robots. Further details on the statistical analysis are reported in the ESI (see Section S.4).

**Figure 3C** demonstrates the benefits of grid based high-throughput research and the self-driving approach over manual experimentation. Within this comparison we assume that systematic manual testing requires the preparation of 500 samples, with a throughput of 20 samples per day, to obtain equivalent information as with HTE of 1,000 samples. Using conventional 1-inch substrates, typical solution concentrations of 30 mg/mL and coating methods, such as spin-coating or blade-coating, a consumption of 188 mg per compound is expected. In contrast, HTE needs only 7.5 mg per compound due to the low concentration of 0.6 mg/mL used in drop-casting. Moreover, even as 288 samples are tested in a single day, HTE guarantees a stable process with highly reproducible data. The substantially reduced number of samples for the self-driving approach with only 60 samples further reduces the amount of consumed materials to 0.45 mg per compound. Following the findings from the virtual robots, the number of samples could even be reduced to 30 (0.225 mg) while maintaining a confidence of over 95% (optimized SDA). Note that for ChemOS the given throughput of 4 samples per day and blend system is based on the iterative degradation process. Nevertheless by parallelization across blend systems, ChemOS is able to run and optimize 16 blend systems simultaneously further accelerating the discovery process.

**Conclusion and Outlook**

We have demonstrated how complex multi-component optimization problems for next generation OSC active layers can be highly accelerated by a novel high throughput film deposition technique combined with automated characterization and machine learning enabled experimental design. Over 2000 different quaternary active layers were tested in 7 days with a materials consumption of less than 15 mg per component. This enables a more thorough stability investigation of novel OPV materials and their interactions with other blend components at ultra-low material consumptions. Implementing a machine learning optimization algorithm, equivalent information about stability minima and maxima could be found with a sample reduction of around 93%. In our specific test case, P3HT and PBQ-QF rich blends show improved stability over PTB7-Th rich blends. O-IDTBR and PCBM can destabilize each other with dramatic consequences for the active layer blend.

As a next step the inclusion of additional characterization methods and target properties e.g. photoluminescence or conductivity could allow to not only optimize for stability but also for electrical performance. Recently, Häse et al. with *Chimera*, demonstrated a hierarchy-based general purpose achievement scalarizing function designed to address multi-parameter optimizations in self-driving laboratories.[30]

**Experimental Section**

*High-throughput coating:*

Stock solutions of the materials in chlorobenzene were manually prepared with a concentration of 0.6 mg/mL. All further ink formulations were mixed by a liquid handling robot [Freedom Evo 100; Tecan Group AG, Switzerland] using 96-well polypropylene microplates [Eppendorf, Germany]. To guarantee good intermixing, an aspirate/dispense step was repeated for three times before drop-casting 25 µL of each mixture onto a pre-patterned glass substrate. The films are dried at 60 °C for 4 minutes. 96 individual films can be formed in 22 min. To obtain stabile morphologies, a thermal annealing step at 120 °C for 15 min was performed. Further descriptions are reported in the supporting information (see Section S.5)

**Authorship contributions**
C.J.B., T.H. and A.A.-G. conceived and supervised the project. S.L. developed the substrate patterning and optimized the robot-process. S.L. and J.D.P. performed the experiments. S.L. evaluated the data and interpreted the experimental results. F.H. and L.M.R designed and implemented adaptations to ChemOS. F.H. and L.M.R designed statistical analyses with the virtual robot, which were implemented and executed by F.H. F.H and L.M.R analyzed and interpreted the results from the virtual robot. T.S. and J.H. set up the robot equipment and supervised the experiments. All authors contributed to the discussion, writing and editing of the manuscript. S.L. and F.H. contributed equally to this work.

**Conflicts of interest**
There are no conflicts to declare.

**Acknowledgments**
The authors want to thank Klaus Burlafinger for fruitful discussions. S.L., T.H. and C.J.B. gratefully thank the Deutsche Forschungsgemeinschaft (DFG) in the framework of SFB 953 "Synthetic Carbon Allotropes" (PN 182849149) for financial support. J.D.P. acknowledges the doctoral fellowship grant of the Colombian Agency COLCIENCIAS. F.H. acknowledges support from the Herchel Smith Graduate Fellowship and the Jacques-Emile Dubois Student Dissertation Fellowship. L.M.R., and A.A-G. were supported by Tata Sons Limited − Alliance Agreement (A32391). F.H., L.M.R., and A.A-G. thank Dr. Anders Frøseth for generous support. A.A-.G. acknowledges the generous support of Natural Resources Canada and the Canada 150 Research Chairs program.

**References**

[1]  Y. Cui, H. Yao, J. Zhang, T. Zhang, Y. Wang, L. Hong, K. Xian, B. Xu, S. Zhang, J. Peng, Z. Wei, F. Gao, J. Hou, *Nat. Commun.* **2019**, *10*, 1.

[2]  X. Liu, Y. Yan, Y. Yao, Z. Liang, *Adv. Funct. Mater.* **2018**, *28*, DOI 10.1002/adfm.201802004.

[3]  K. Li, Y. Wu, Y. Tang, M. Pan, W. Ma, H. Fu, C. Zhan, J. Yao, *Adv. Energy Mater.* **2019**, *1901728*, 1901728.

[4]  R. A. Potyrailo, B. J. Chisholm, W. G. Morris, J. N. Cawse, W. P. Flanagan, L. Hassib, C. A. Molaison, K. Ezbiansky, G. Medford, H. Reitz, *J. Comb. Chem.* **2003**, *5*, 472.

[5]  D. G. Anderson, S. Levenberg, R. Langer, *Nat. Biotechnol.* **2004**, *22*, 863.

[6]  A. Jaklenec, A. C. Anselmo, J. Hong, A. J. Vegas, M. Kozminsky, R. Langer, P. T. Hammond, D. G. Anderson, *ACS Appl. Mater. Interfaces* **2016**, *8*, 2255.

[7] X. Rodríguez-Martínez, A. Sánchez-Díaz, G. Liu, M. A. Niño, J. Cabanillas-Gonzalez, M. Campoy-Quiles, *Org. Electron. physics, Mater. Appl.* **2018**, *59*, 288.

[8] A. Sánchez-Díaz, X. Rodríguez-Martínez, L. Córcoles-Guija, G. Mora-Martín, M. Campoy-Quiles, *Adv. Electron. Mater.* **2018**, *4*, DOI 10.1002/aelm.201700477.

[9] A. Teichler, J. Perelaer, U. S. Schubert, *Mater. Res. Soc. Symp. Proc.* **2012**, *1390*, 103.

[10] D. P. Tabor, L. M. Roch, S. K. Saikin, C. Kreisbeck, D. Sheberla, J. H. Montoya, S. Dwaraknath, M. Aykol, C. Ortiz, H. Tribukait, C. Amador-Bedolla, C. J. Brabec, B. Maruyama, K. A. Persson, A. Aspuru-Guzik, *Nat. Rev. Mater.* **2018**, *3*, 5.

[11] J. P. Correa-Baena, K. Hippalgaonkar, J. van Duren, S. Jaffer, V. R. Chandrasekhar, V. Stevanovic, C. Wadia, S. Guha, T. Buonassisi, *Joule* **2018**, *2*, 1410.

[12] A. Aspuru-Guzik, K. Persson, *Mission Innov. Innov. Chall. 6* **2018**.

[13] K. T. Butler, D. W. Davies, H. Cartwright, O. Isayev, A. Walsh, *Nature* **2018**, *559*, 547.

[14] H. Winicov, J. Schainbaum, J. Buckley, G. Longino, J. Hill, C. E. Berkoff, *Anal. Chim. Acta* **1978**, *103*, 469.

[15] P. Nikolaev, D. Hooper, F. Webber, R. Rao, K. Decker, M. Krein, J. Poleski, R. Barto, B. Maruyama, *npj Comput. Mater.* **2016**, *2*, DOI 10.1038/npjcompumats.2016.31.

[16] D. Xue, P. V. Balachandran, J. Hogden, J. Theiler, D. Xue, T. Lookman, *Nat. Commun.* **2016**, *7*, 1.

[17] J. M. Granda, L. Donina, V. Dragone, D. L. Long, L. Cronin, *Nature* **2018**, *559*, 377.

[18] P. J. Kitson, G. Marie, J.-P. Francoia, S. S. Zalesskiy, R. C. Sigerson, J. S. Mathieson, L. Cronin, *Science (80-. ).* **2018**, *359*, 314.

[19] B. P. MacLeod, F. G. L. Parlane, T. D. Morrissey, F. Häse, L. M. Roch, K. E. Dettelbach, R. Moreira, L. P. E. Yunker, M. B. Rooney, J. R. Deeth, V. Lai, G. J. Ng, H. Situ, R. H. Zhang, A. Aspuru-Guzik, J. E. Hein, C. P. Berlinguette, *arXiv* **2019**, arXiv:1906.05398.

[20] X. Du, T. Heumueller, W. Gruber, A. Classen, T. Unruh, N. Li, C. J. Brabec, *Joule* **2019**, *3*, 215.

[21] A. Distler, P. Kutka, T. Sauermann, H. J. Egelhaaf, D. M. Guldi, D. Di Nuzzo, S. C. J. Meskers, R. A. J. Janssen, *Chem. Mater.* **2012**, *24*, 4397.

[22] E. T. Hoke, I. T. Sachs-Quintana, M. T. Lloyd, I. Kauvar, W. R. Mateker, A. M. Nardes, C. H. Peters, N. Kopidakis, M. D. McGehee, *Adv. Energy Mater.* **2012**, *2*, 1351.

[23] M. Salvador, N. Gasparini, J. D. Perea, S. H. Paleti, A. Distler, L. N. Inasaridze, P. A. Troshin, L. Lüer, H. J. Egelhaaf, C. Brabec, *Energy Environ. Sci.* **2017**, *10*, 2005.

[24] C. Zhang, T. Heumueller, S. Leon, W. Gruber, K. Burlafinger, X. Tang, J. D. Perea, I. Wabra, A. Hirsch, T. Unruh, N. Li, C. J. Brabec, *Energy Environ. Sci.* **2019**, *12*, 1078.

[25] S. Chen, Y. Hou, H. Chen, X. Tang, S. Langner, N. Li, T. Stubhan, I. Levchuk, E. Gu, A. Osvet, C. J. Brabec, *Adv. Energy Mater.* **2018**, *8*, 1.

[26] C. Xie, X. Tang, M. Berlinghof, S. Langner, S. Chen, A. Späth, N. Li, R. H. Fink, T. Unruh, C. J. Brabec, *ACS Appl. Mater. Interfaces* **2018**, *10*, 23225.

[27] L. M. Roch, F. Häse, C. Kreisbeck, T. Tamayo-Mendoza, L. P. E. Yunker, J. E. Hein, A. Aspuru-Guzik, *Chemrxiv* **2018**, 10.26434/chemrxiv.5952655.

[28] L. M. Roch, F. Häse, C. Kreisbeck, T. Tamayo-Mendoza, L. P. E. Yunker, J. E. Hein,

A. Aspuru-Guzik, *Sci. Robot.* **2018**, *3*, eaat5559.

[29] F. Häse, L. M. Roch, C. Kreisbeck, A. Aspuru-Guzik, *ACS Cent. Sci.* **2018**, *4*, 1134.

[30] F. Häse, L. M. Roch, A. Aspuru-Guzik, *Chem. Sci.* **2018**, *9*, 7642.

[31] Https://github.com/aspuru-guzik-group/quaterny_opvs, **2019**.

**TOC:**

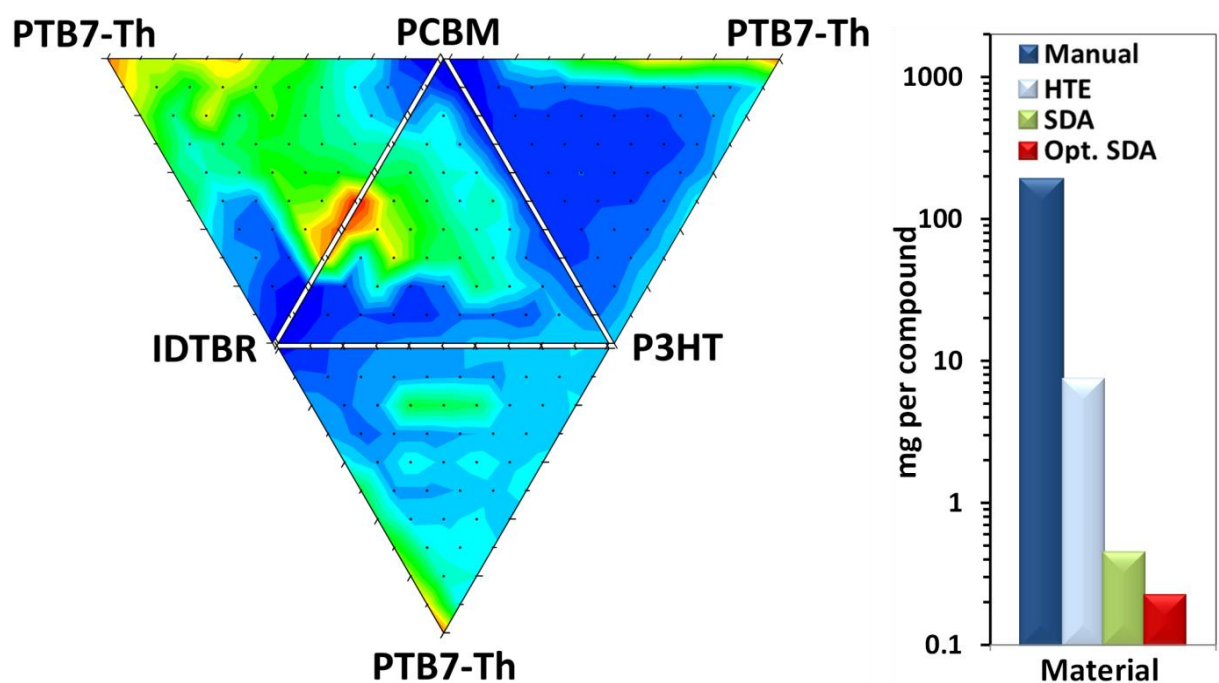

# Supporting Information

## S.1 Experiment planning with robotics constraints

The robotic system for blend formulation has a finite dispensing accuracy of approximately 1 µL (see main text). As such, the blend compositions realized by the automated platform can differ from the blend formulations suggested by ChemOS. Specifically, the robotic system could, by chance, prepare two polymer blends with almost identical compositions, which ultimately results in the repetition of a previous experiment. Here, we aim to accelerate the discovery of photostable polymer blends. Hence, redundant experiments need to be avoided and a more uniform coverage of the search space is required. To this end, we constructed a high-resolution grid of possible parameter points where any two blend compositions can be realized distinguishably by the automated platform despite the finite dispensing accuracy. Given the dispensing accuracy of 2 wt.-% (1 µL), the high-resolution grid was constructed with 19,317 different parameter points.

We extended ChemOS by an additional module to transform continuous experimental parameters suggested by the experiment planner to the grid of distinguishable experiments which the automated platform can realize. Note that ChemOS' modular architecture enables the simplified integration of this extra step in the self-driving approach without interfering with existing modules. This additional module extends an iteration of the self-driving approach as follows. (i) ChemOS first calls the experiment planner, which was chosen to be Phoenics for this application, to receive a set of continuous experimental parameters suggested based on previous measurements. (ii) Then, the suggested parameters are projected onto the grid of distinguishable experiments, using the Euclidean metric. (iii) Finally, the identified closest grid point is suggested as the next experiment and sent to the robotic system. Step (ii) constitutes an addition to the conventional self-driving approach. Note, that the experiment planner is still updated with measurements associated to the experimental conditions it proposed, as opposed to the grid parameters which were realized. This policy ensures that the robotic space and the application space on which the experiment planner operates are well defined and kept separate.

In the self-driving approach, we employed two different merit functions in two different executions to quantitatively rank individual blend compositions. While in the first execution the obtained photodegradation measurement served directly as the merit, the second execution targeted a minimization of the logarithm of the photodegradation measurement.

## S.2 Results of high-throughput experimentation and self-driving laboratory

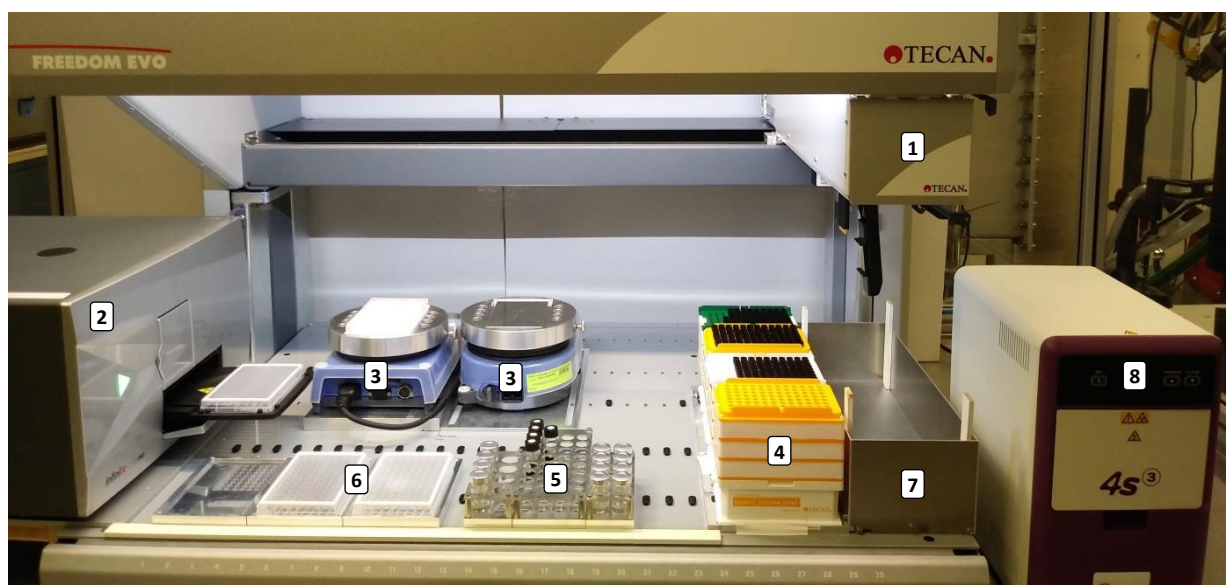

*Figure S1.* Front view of our semi-automatic robot system: *(1)* robot arm with four pipetting channels up to 1 mL each; *(2)* spectrometer with Abs and PL mode; *(3)* two hotplates; *(4)* different sizes of tips; *(5)* stock solutions for experiment; *(6)* 96-well microplates as experiment-vessels; *(7)* waste container for tips; *(8)* heat sealer to fuse microplates with aluminum foils.

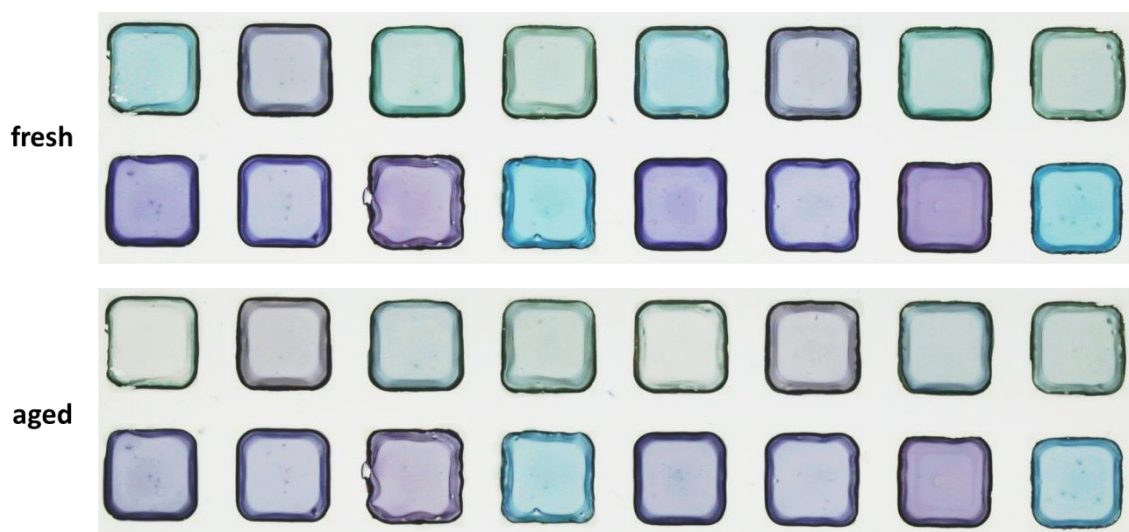

*Figure S3A.* Fresh films compared to aged films after 18 hours of continuous illumination.

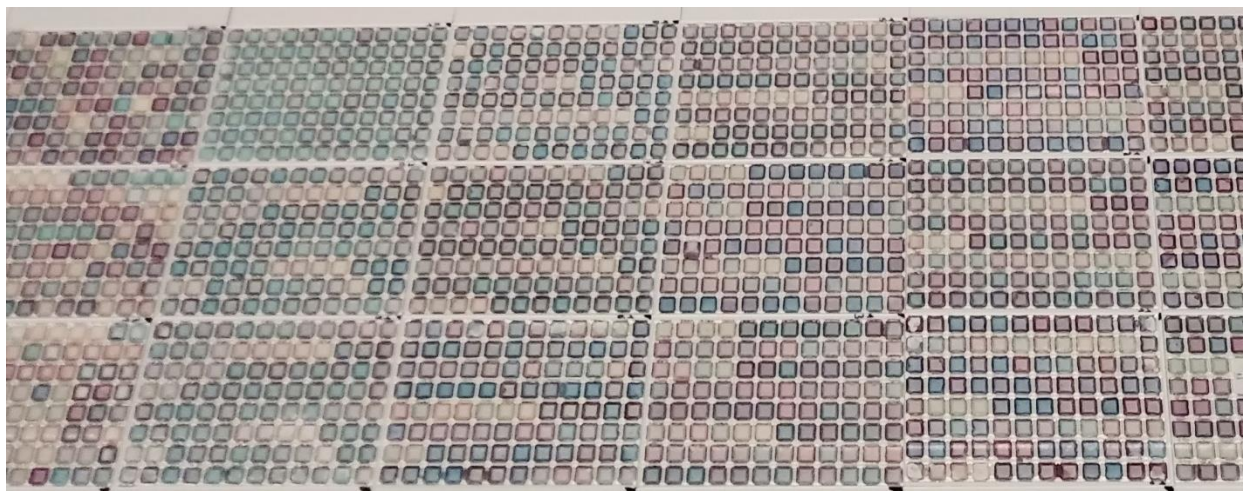

*Figure S3B. Overview of a subset from all the samples prepared and characterized in this study.*

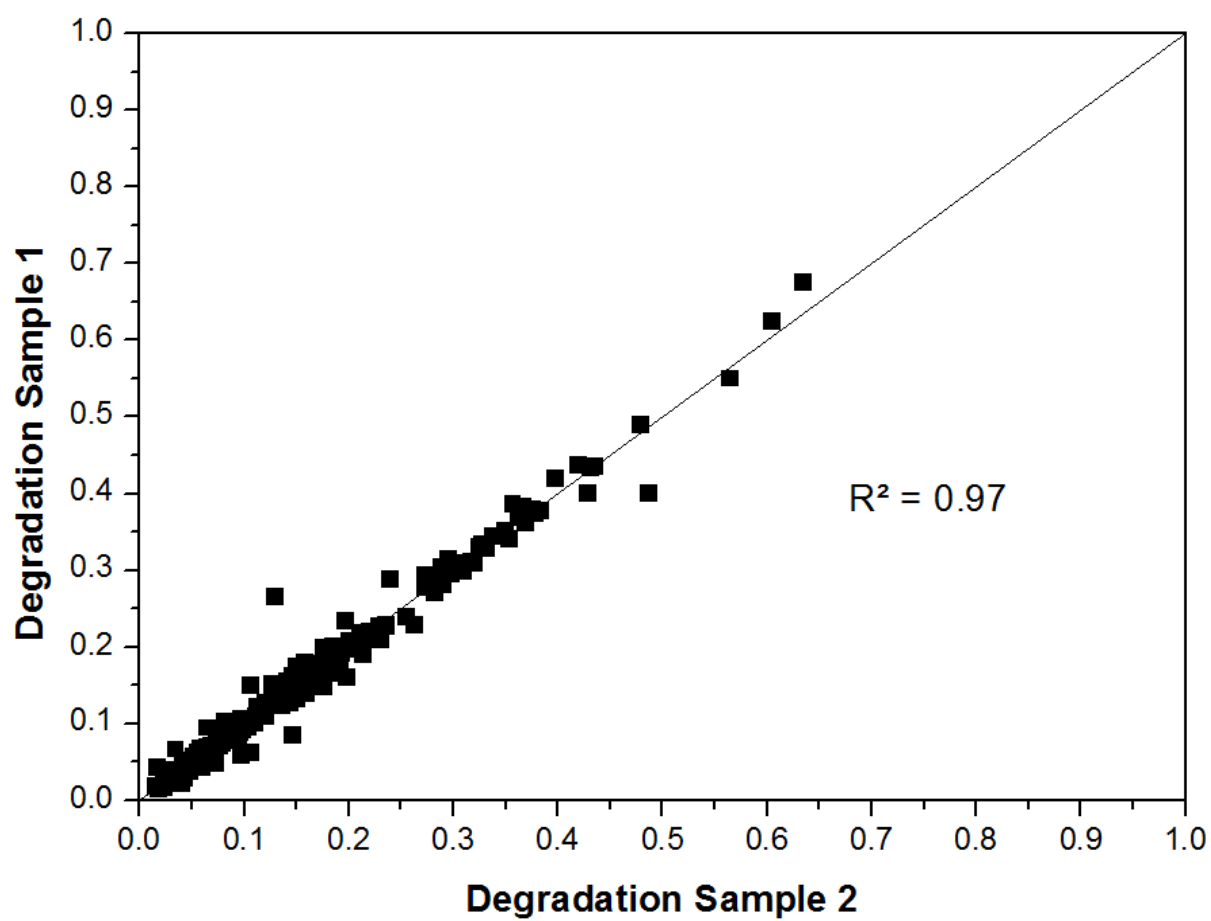

*Figure S3. Reproducibility of the high-throughput drop-casting film formation.*

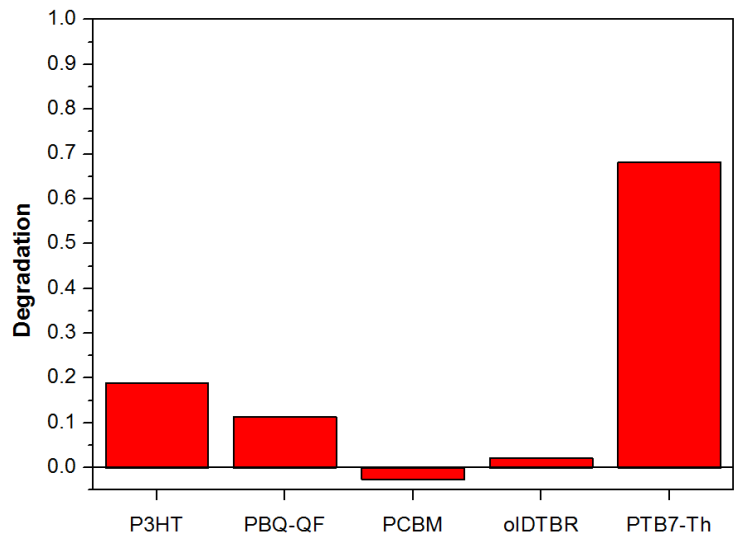

*Figure S4.* *Stability of the materials used in this study. PCBM has a negative value due to slight increase of the spectra.*

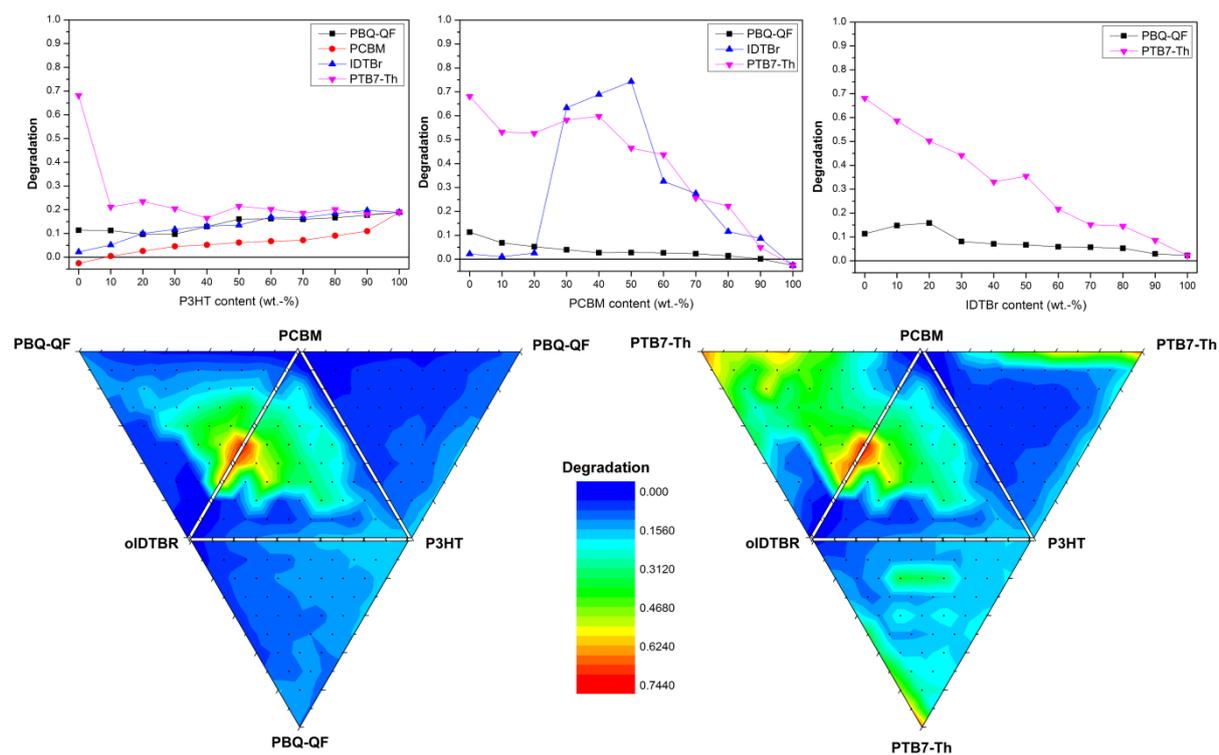

*Figure S5.* *Stability graphs of binary and ternary mixtures for PBQ-QF based system (left) and PTB7-Th based system (right).*

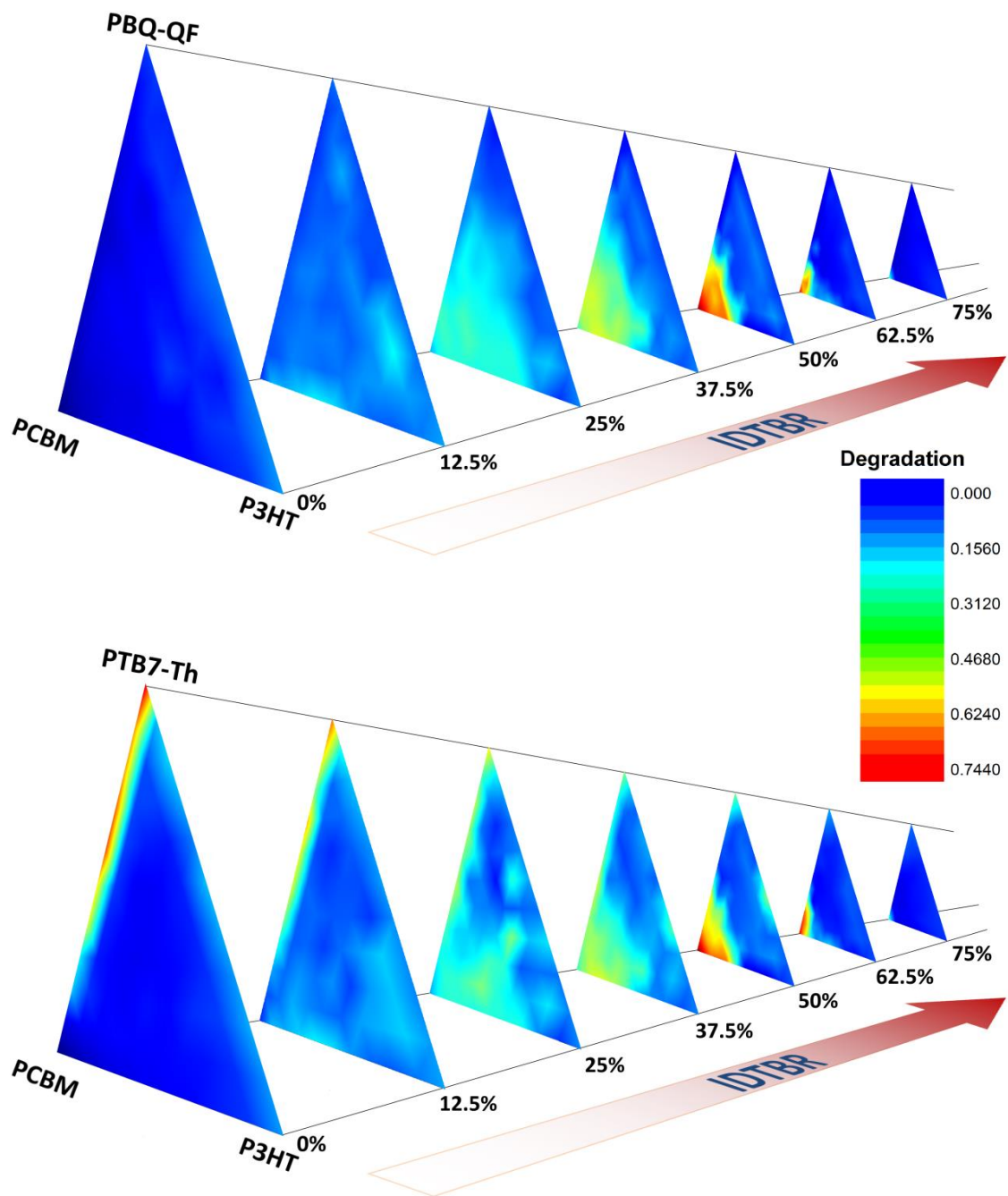

*Figure S6.* Stability graphs of quaternary mixtures for PBQ-QF based system (top) and PTB7-Th based system (bottom).

*Table S1.* The 10 best compositions found by the self-driving approach and grid-HTE.

| PBQ-QF | P3HT wt.-fraction | WF3 wt.-fraction | PCBM wt.-fraction | IDTBr wt.-fraction | Degradation arb. unit | PTB7-Th | P3HT wt.-fraction | PCE10 wt.-fraction | PCBM wt.-fraction | IDTBr wt.-fraction | Degradation arb. unit |
|---|---|---|---|---|---|---|---|---|---|---|---|
| ChemOS Merit Function 1 | 0 | 0.5 | 0.5 | 0 | 0.0165 | ChemOS Merit Function 1 | 0.14 | 0.04 | 0.2 | 0.62 | 0.0298 |
| | 0 | 0.56 | 0.44 | 0 | 0.0171 | | 0.3 | 0.48 | 0.22 | 0 | 0.0506 |
| | 0 | 0.52 | 0.48 | 0 | 0.0202 | | 0.16 | 0.46 | 0.28 | 0.1 | 0.0558 |
| | 0 | 0.6 | 0.4 | 0 | 0.0217 | | 0.24 | 0.5 | 0.26 | 0 | 0.0610 |
| | 0 | 0.58 | 0.42 | 0 | 0.0296 | | 0.58 | 0.08 | 0.34 | 0 | 0.0638 |
| | 0 | 0.68 | 0.32 | 0 | 0.0303 | | 0.24 | 0.28 | 0.42 | 0.06 | 0.0668 |
| | 0 | 0.62 | 0.38 | 0 | 0.0316 | | 0.26 | 0.56 | 0.18 | 0 | 0.0670 |
| | 0.18 | 0.46 | 0.36 | 0 | 0.0356 | | 0.5 | 0 | 0 | 0.5 | 0.0697 |
| | 0.56 | 0 | 0.12 | 0.32 | 0.0362 | | 0.26 | 0.56 | 0.1 | 0.08 | 0.0697 |
| | 0 | 0.7 | 0.3 | 0 | 0.0430 | | 0.3 | 0.46 | 0.24 | 0 | 0.0700 |
| ChemOS Merit Function 2 | 0.26 | 0.1 | 0.62 | 0.02 | 0.0296 | ChemOS Merit Function 2 | 0.06 | 0.12 | 0.2 | 0.62 | 0.0315 |
| | 0.4 | 0.1 | 0.12 | 0.38 | 0.0379 | | 0.06 | 0.12 | 0.22 | 0.6 | 0.0362 |
| | 0.34 | 0.34 | 0.3 | 0.02 | 0.0400 | | 0.12 | 0.06 | 0.18 | 0.64 | 0.0385 |
| | 0.26 | 0.34 | 0.08 | 0.32 | 0.0432 | | 0.08 | 0.1 | 0.2 | 0.62 | 0.0525 |
| | 0.44 | 0.06 | 0.04 | 0.46 | 0.0448 | | 0.08 | 0.1 | 0.18 | 0.64 | 0.0540 |
| | 0.36 | 0.24 | 0.04 | 0.36 | 0.0453 | | 0.1 | 0.08 | 0.2 | 0.62 | 0.0579 |
| | 0.4 | 0.1 | 0.1 | 0.4 | 0.0472 | | 0.22 | 0.38 | 0.38 | 0.02 | 0.0624 |
| | 0.24 | 0.36 | 0.38 | 0.02 | 0.0480 | | 0.38 | 0.28 | 0.34 | 0 | 0.0628 |
| | 0.3 | 0.1 | 0.6 | 0 | 0.0568 | | 0.18 | 0.4 | 0.08 | 0.34 | 0.0642 |
| | 0.24 | 0.1 | 0.66 | 0 | 0.0599 | | 0.08 | 0.12 | 0.18 | 0.62 | 0.0672 |
| Grid-HTE | 0 | 0.2 | 0.8 | 0 | 0.0146 | Grid-HTE | 0.2 | 0 | 0.8 | 0 | 0.0257 |
| | 0 | 0.3 | 0.7 | 0 | 0.0227 | | 0.16 | 0.52 | 0.08 | 0.24 | 0.0397 |
| | 0.1 | 0.1 | 0.8 | 0 | 0.0248 | | 0.24 | 0.38 | 0.14 | 0.24 | 0.0453 |
| | 0.2 | 0 | 0.8 | 0 | 0.0257 | | 0.3 | 0 | 0.7 | 0 | 0.0455 |
| | 0 | 0.4 | 0.6 | 0 | 0.0265 | | 0.4 | 0 | 0.6 | 0 | 0.0519 |
| | 0 | 0.6 | 0.4 | 0 | 0.0273 | | 0.04 | 0.16 | 0.78 | 0.02 | 0.0539 |
| | 0 | 0.5 | 0.5 | 0 | 0.0279 | | 0.02 | 0.2 | 0.18 | 0.6 | 0.0547 |
| | 0.02 | 0.18 | 0.24 | 0.56 | 0.0333 | | 0.3 | 0 | 0.1 | 0.6 | 0.0583 |
| | 0.2 | 0.1 | 0.7 | 0 | 0.0359 | | 0.24 | 0.04 | 0.7 | 0.02 | 0.0584 |
| | 0 | 0.2 | 0.1 | 0.7 | 0.0380 | | 0.3 | 0 | 0.2 | 0.5 | 0.0584 |

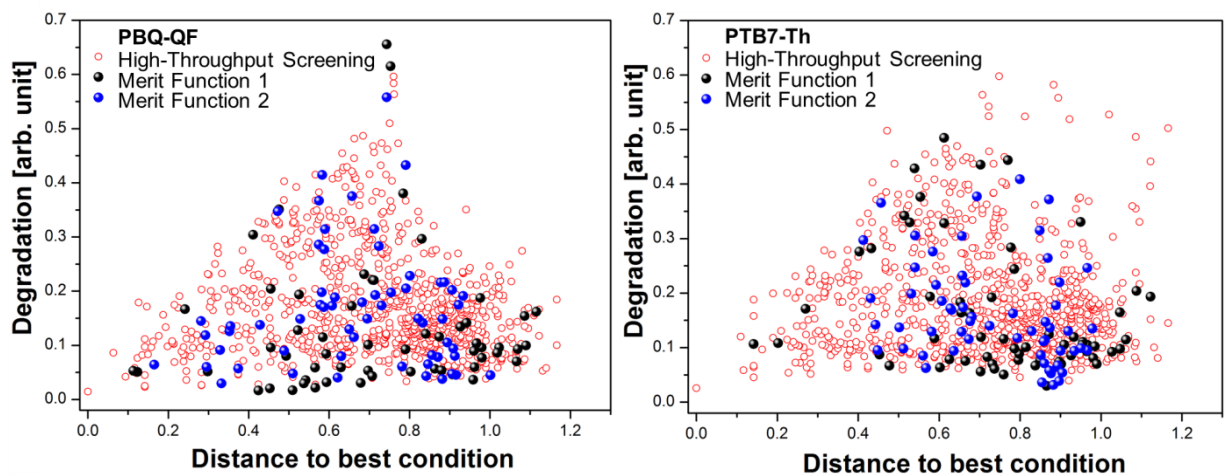

*Figure S7.* Comparison between grid-based high-throughput experimentation and ChemOS optimization for the PBQ-QF based system. Here each data point is distance related to the best found HTE composition ($distance = \sqrt{\Delta x^2 + \Delta y^2 + \Delta z^2 + \Delta t^2}$).

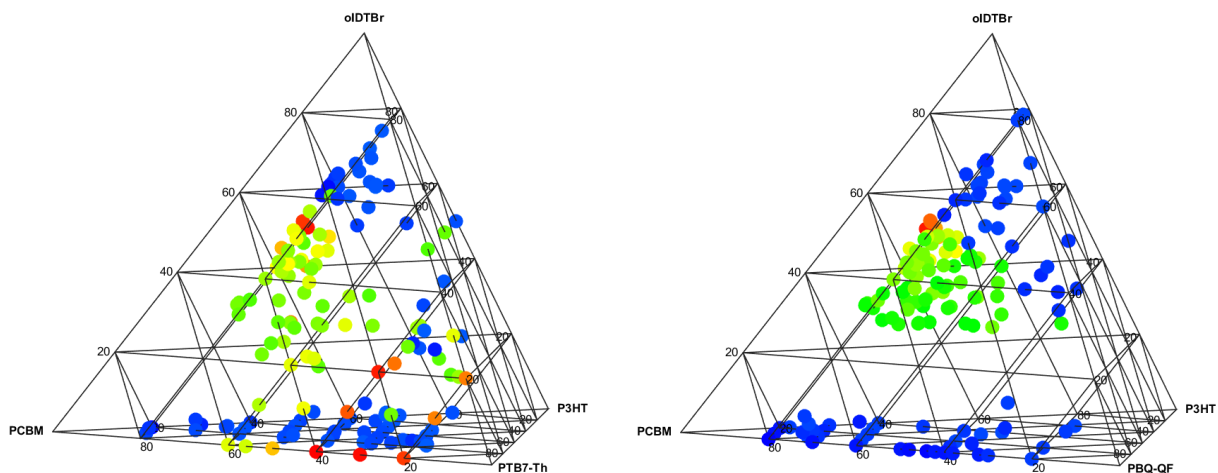

*Figure S8.* Tetrahedral plot of the most stable (blue points) and most unstable (green/red points) compositions. The dataset includes data points from both strategies, HTE and SDA.

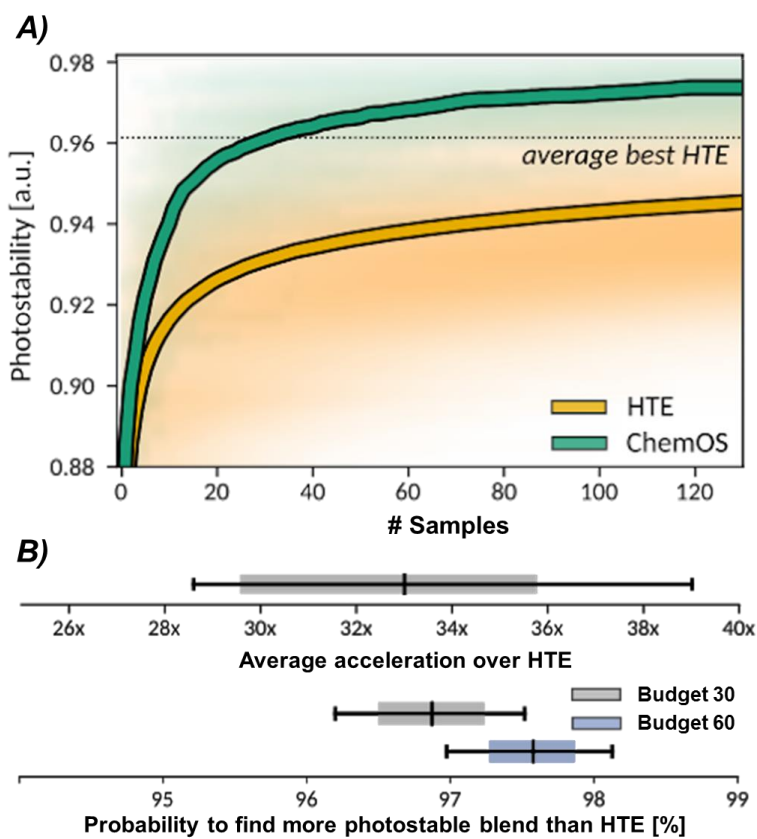

*Figure S9.* A) Performances of high-throughput experimentation (red traces) and ChemOS/Phoenics (blue traces) on the virtual robots related to the PBQ-QF based system. Note that the traces were averaged over 10,000 independent grid samplings and 200 independent ChemOS optimizations. B) Top: Statistical output of the virtual robot showing the acceleration of SDA over HTE; Bottom: Confidence to improve on HTE within given budget of 30 or 60 samples. Both are related to the PBQ-QF based system.

## S.3 Construction of virtual robots

The statistical significance of the experimentally obtained performance difference between the self-driving approach and high-throughput experimentation (HTE) was assessed in detail with the construction of virtual robots. Virtual robots can be designed *via* probabilistic machine learning (ML) models to emulate an experimental procedure *in silico*.[1] Assuming that the virtual robot reproduces the measurements of previous experiments sufficiently well, the merit of new experiments can be estimated computationally without conducting further experiments. Consequently, virtual robots can serve as a benchmark to determine performance differences between different experiment planning strategies.

We construct virtual robots from Bayesian neural networks (BNNs) which are trained to reproduce the photo-degradation measured for individual polymer blends in the HTE approach. BNNs constitute probabilistic ML models which can be trained *via* variational expectation maximization in a Bayesian framework. Thus, BNNs are intrinsically robust to overfitting and can infer both the expected photo-degradation and the degree of experimental noise to resemble experimental conditions. The applicability of BNNs to emulate experiments has already been demonstrated in the context of modelling high-performance liquid chromatography signals.[1]

A total of 850 experiments (81.7%) were randomly selected from the 1,041 HTE experiments for both blend systems to train BNNs with 5-fold cross-validation. The remaining 190 experiments (18.3%) were used as a test set to determine the generalization performance of the trained BNNs. BNNs were constructed as fully connected feedforward networks with three hidden layers. To account for the fact that the photo-degradation of any given polymer blend cannot be negative, we selected ReLU activation functions for the output layer. Weights and biases were modelled with Gaussian priors. Additional network hyperparameters and the selected values are reported in **Table S2**.

**Table S2**: *Hyperparameters for the Bayesian neural networks used as virtual robots to emulate the photo-degradation of individual polymer blends.*

| Hyperparameter | Selected value |
| --- | --- |
| Number of hidden layers | 3 |
| Neurons per hidden layer | 120 |
| Hidden layer activations | Leaky ReLU |
| Learning rate | 0.001 |
| Dropout rate | 0.2 |

Both polymer blend ratios and associated photo-degradations were rescaled prior to being presented to the BNNs. Polymer blend ratios were transformed from the 4-simplex, $\Delta^4$, to the three-dimensional unit cube, while individual photo-degradations were divided by the average photo-degradation taken over the 850 experiments used for cross-validation.

Prediction accuracies for the BNNs trained on the two blend systems are illustrated in **Figure S10** We observe that the BNNs are capable of reproducing the experimental measurements of the test sets with coefficients of determination of 0.88 (0.87) on the PBQ-QF (PTB7-Th) blend system. The agreement of coefficients of determination between the test and the training sets indicates good transferability of the BNNs.

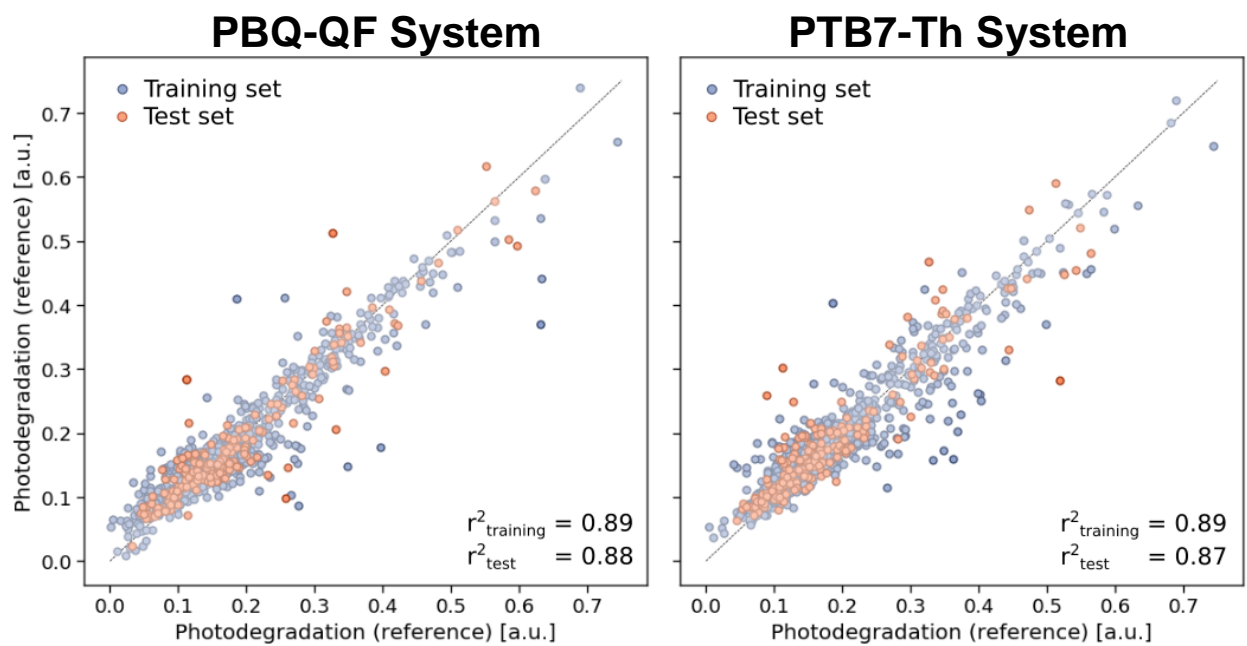

**Figure S10:** *Prediction accuracies of the virtual robots constructed from Bayesian neural networks on the two studied blend systems. Predicted photo-degradations are compared to measurements obtained from high-throughput experimentation.*

## S.4  Statistical analysis of virtual robot optimizations

The expected benefits of the self-driving approach over conventional HTE are determined from simulations on the virtual robot constructed in Sec. S.3.

In a first comparison, we estimate the number of experiments required for ChemOS to discover a polymer blend which is at least as photostable as the average most photostable blend discovered with HTE (884 experiments in total, averaged over 10,000 independent repetitions). The average number of experiments is computed from the 200 independent ChemOS optimization traces, and confidence bands (68% and 95%) are estimated with bootstrapping. In addition, we assess the expected performance of a single optimization run by bootstrapping confidence bands (68% and 95%) for a single trace. Results are illustrated in the manuscript (see **Figure 3**). Note that ChemOS required fewer experiments to locate polymer blends which are more photostable than the average most photostable blend identified in HTE in all 200 iterations.

Based on the estimated number of experiments, we further compute the acceleration of the experimentation campaign, relating the number of HTE experiments to the number of ChemOS experiments. Confidence bands are calculated as before. Further, we estimate the probability of ChemOS having identified a more photostable blend than HTE for a given budget. This probability is computed based on the number of times a ChemOS trace located a more photostable blend than a given HTE trace after a certain number of experiments. The probability is then averaged over all HTE traces.

## S.5  Experimental

*Materials:*

Chlorobenzene (99%) was purchased from Alfa Aesar [USA]. oIDTBR and PTB7-Th were ordered from 1-material [USA]. P3HT and PCBM were provided by Merck KGaA [Germany] and Solenne BV [Netherlands], respectively. PBQ-QF was purchased from Advent Technologies Inc. [Greece].

*Multi-well substrates:*

The patterning of 96 wells on top of a glass substrate (125x85 mm²) is performed with a dispensing robot [I&J 4100-LF, Fisnar Inc., USA] using an UV-curable adhesive [DELO LP 655, DELO Industrie Klebstoffe GmbH & Co. KGaA, Germany], which is cured for 5 min under UV light [UVACUBE 100, Dr. Hönle AG, Germany].

*Characterization and degradation:*

Absorbance measurements (300-800 nm, 5 nm increment) were performed with a microplate reader [Infinity m200-Pro; Tecan Group AG, Switzerland]. The samples were then degraded under ambient conditions with a metal-halide lamp of one sun intensity for 18 hours before the absorbance spectra were recorded again. The absorbance loss was calculated by subtracting the aged from the fresh spectrum and integrating the resulting differences across all wavelengths.

**References**

[1]  F. Häse, L. M. Roch, A. Aspuru-Guzik, *Chem. Sci.* **2018**, *9*, 7642.